\documentclass[conference]{ieeeconf}
\IEEEoverridecommandlockouts
\usepackage{acro}
\let\labelindent\relax
\usepackage{enumitem}

\usepackage{graphics}
\usepackage{graphicx} % For including and formatting image files
\usepackage{multirow}
\usepackage{longtable}
\usepackage{epsfig}
\usepackage{epstopdf}

\usepackage{amsmath, mathrsfs, dsfont}
\usepackage{amssymb}
\usepackage{amsfonts}
\usepackage{bm}

\usepackage{cite}
\usepackage{verbatim}

\usepackage{amscd}
\usepackage{color}
\definecolor{lgray}{gray}{0.6}
\usepackage{rotating}

\usepackage[font=footnotesize]{subcaption}

\usepackage{mathptmx}
\usepackage[11pt]{moresize}
\usepackage{flushend}
\usepackage[stable]{footmisc}

\usepackage{algorithmicx}
\usepackage{algorithm}
\usepackage{algpascal}
\usepackage{float}
\usepackage{algc}
\usepackage{algcompatible}
\usepackage{algpseudocode}

\usepackage[vcentermath]{youngtab} % For typesetting Young Tableaux
\usepackage{url}
\newcommand{\Bolda}				{ \mathbf{a} }

\newcommand{\Boldb}				{ \mathbf{b} }

\newcommand{\Bolde}				{ \mathbf{e} }

\newcommand{\BoldH}				{ \mathbf{H} }
\newcommand{\Boldh}				{ \mathbf{h} }
\newcommand{\BoldI}				{ \mathbf{I} }

\newcommand{\BoldJ}				{ \mathbf{J} }

\newcommand{\BoldP}				{ \mathbf{P} }
\newcommand{\Boldp}				{ \mathbf{p} }
\newcommand{\BoldQ}				{ \mathbf{Q} }

\newcommand{\BoldR}				{ \mathbf{R} }

\newcommand{\Boldr}				{ \mathbf{r} }

\newcommand{\BoldT}				{ \mathbf{T} }

\newcommand{\Boldv}				{ \mathbf{v} }
\newcommand{\Boldw}				{ \mathbf{w} }

\newcommand{\Boldx}				{ \mathbf{x} }

\newcommand{\Boldy}				{ \mathbf{y} }

\newcommand{\Bolddt}			{ \mathbf{dt} }

\newcommand{\0}					{ \boldsymbol{0} }
\newcommand{\1}					{ \boldsymbol{1} }

\newcommand{\BoldPhi}			{ \boldsymbol{\Phi} }

\newcommand{\BoldSigma}			{ \boldsymbol{\Sigma} }

\newcommand{\Boldeps}			{ \boldsymbol{\epsilon} }

\newcommand{\BoldPsi}           {\boldsymbol{\Psi}}

\newcommand\T{\rule{0pt}{2.6ex}}       % Top strut
\newcounter{inlineenum}
\renewcommand{\theinlineenum}{\alph{inlineenum}}

\newcommand{\brmrk}[1]{\begin{remark} \label{#1} }
	\newcommand{\ermrk}{ \hfill $\bigtriangleup$    \end{remark} \vspace{1mm} }

\newtheorem{exercise}{Exercise}[section]
\newcommand{\boex}[1]{\begin{example} \label{#1} --- \rm}
	\newcommand{\eoex}{ \hfill $\bigtriangleup$    \end{example} \vspace{1mm} }
\newtheorem{example}{Example}[section]
\newcommand{\bohw}[1]{\begin{exercise} \label{#1} -- \rm}
	\newcommand{\eohw}{ \hfill    \end{exercise} \vspace{1mm} }
\newtheorem{assumption}{Assumption}[section]
\newcommand{\boass}[1]{\begin{assumption} \label{#1} -- \rm}
	\newcommand{\eoass}{ \hfill    \end{assumption} \vspace{1mm} }

\renewcommand{\baselinestretch}{1.0}

\newcommand{\black}{\color{black}}

\newcommand{\red}{\color{red}}

\definecolor{brinkpink}{rgb}{1.00, 0.33, 0.64}



\usepackage{tikz}
\usepackage{wrapfig}
\usepackage{stfloats}
\DeclareMathAlphabet\mathbfcal{OMS}{cmsy}{b}{n}
\DeclareAcronym{APC}{
	short=APC,
	long=Antenna Phase Center,
}

\DeclareAcronym{COM}{
short=COM,
long=Center of Mass,
}

\DeclareAcronym{CV}{
short=CV,
long=Connected Vehicle,
}

\DeclareAcronym{CAV}{
	short=CAV,
	long=Connected Automated Vehicles,
}

\DeclareAcronym{DCB}{
short=DCB,
long=Differential Code Bias,
}

\DeclareAcronym{DSRC}{
	short=DSRC,
	long=Dedicated Short-Range Communications,
}

\DeclareAcronym{CDMA}{
	short=CDMA,
	long=Code Division Multiple Access,
}

\DeclareAcronym{CDF}{
	short=CDF,
	long=Cumulative Distribution Function,
}

\DeclareAcronym{CE-CERT}{
	short=CE-CERT,
	long=College of Engineering Center for Environmental Research and Technology,
}

\DeclareAcronym{CME}{
	short=CME,
	long=common-mode errors,
}

\DeclareAcronym{FDMA}{
	short=FDMA,
	long=Frequency Division Multiple Access,
}
\DeclareAcronym{DGPS}{
	short=DGPS,
	long=Differential Global Positioning System,
}
\DeclareAcronym{DLR}{
	short=DLR,
	long=German Aerospace Center,
}
\DeclareAcronym{DGNSS}{
short=DGNSS,
long=Differential GNSS,
}
\DeclareAcronym{ECEF}{
short=ECEF,
long=Earth-Centered Earth-Fixed,
}
\DeclareAcronym{ECI}{
	short=ECI,
	long=Earth-Centered Inertial,
}
\DeclareAcronym{EKF}{
	short=EKF,
	long=Extended Kalman Filter,
}
\DeclareAcronym{GPS}{
	short=GPS,
	long=Global Positioning System,
}
\DeclareAcronym{GNSS}{
short=GNSS,
long=Global Navigation Satellite Systems,
}
\DeclareAcronym{GPSSPS}{
short=GPS SPS,
long=GPS standard positioning service,
}

\DeclareAcronym{HMM}
{short = HMM, long = Hidden Markov Model}

\DeclareAcronym{HD}
{short = HD, long = Horizontal Distance}

\DeclareAcronym{HiD}{
	short=Hi-Def,
	long=High-definition,
}

\DeclareAcronym{IMU}{
short=IMU,
long=Inertial Measurement Unit,
}

\DeclareAcronym{ILP}{
short=ILP,
long=Integer Linear Programming,
}

\DeclareAcronym{I2V}{
short=I2V,
long=Infrastructure-to-Vehicle,
}
\DeclareAcronym{IGS}{
	short=IGS,
	long=International GNSS Service,
}
\DeclareAcronym{ISB}{
	short=ISB,
	long=inter-system bias,
}

\DeclareAcronym{LLMM}
{short = LLMM, long = Lane-level Map-matching}

\DeclareAcronym{LD}
{short = LD, long = Lane Determination}

\DeclareAcronym{LOS}
{short = LOS, long = line-of-sight}

\DeclareAcronym{RLMM}
{short = RLMM, long = Road-level Map-matching}

\DeclareAcronym{MSE}
{short = MSE, long = Mean Square Error}

\DeclareAcronym{MGEX}
{short = MGEX, long = Multi-GNSS Experiment}

\DeclareAcronym{MAP}
{short = MAP, long = maximum-a-posteriori}
\DeclareAcronym{OS}{
	short=OS,
	long=Open Service,
}

\DeclareAcronym{OSB}{
short=OSB,
long=Observable-specific Code Biases,
}
\DeclareAcronym{OSR}{
short=OSR,
long=Observation Space Representation,
}
\DeclareAcronym{PPP}{
	short=PPP,
	long=Precise Point Positioning,
} 

\DeclareAcronym{RTPPP}{
	short=RT-PPP,
	long=Real-time PPP,
}

\DeclareAcronym{PPP-AR}{
short=PPP-AR,
long=Precise Point Positioning Ambiguity Resolution,
} 

\DeclareAcronym{PP}{
short=PP,
long=Post Processing,
}

\DeclareAcronym{PVA}{
	short=PVA,
	long={position, velocity, acceleration},
} 

\DeclareAcronym{RTCM}{
short=RTCM,
long=Radio Technical Commission for Maritime Services,
}
\DeclareAcronym{RTK}{
short=RTK,
long=Real-time Kinematic Positioning,
}
\DeclareAcronym{SBAS}{
short=SBAS,
long=Satellite Based Augmentation Systems,
}

\DeclareAcronym{SNR}{
short=SNR,
long=Signal-to-Noise Ratio,
}
\DeclareAcronym{SSR}{
short=SSR,
long=State Space Representation,
}
\DeclareAcronym{SPS}{
	short=SPS,
	long=Standard Positioning Service,
}

\DeclareAcronym{SAE}{
	short=SAE,
	long=Society of Automotive Engineers,
}
\DeclareAcronym{STEC}{
	short=STEC,
	long=Slant Total Electron Content,
}
\DeclareAcronym{VTEC}{
	short=VTEC,
	long=Vertical Total Electron Content,
}

\DeclareAcronym{SH}{
	short=SH,
	long=Spherical Harmonic,
}
\DeclareAcronym{TGD}{
short=TGD,
long=Timing Goup Delay,
}
\DeclareAcronym{TD}{
	short=TD,
	long=Threshold Decisions,
}
\DeclareAcronym{TOP}{
	short=TOP,
	long=time-of-signal-propagation,
}
\DeclareAcronym{TOT}{
	short=TOT,
	long=time-of-signal-transmission,
}
\DeclareAcronym{TOR}{
	short=TOR,
	long=time-of-signal-reception,
}
\DeclareAcronym{ZTD}{
short=ZTD,
long=Zenith Troposphere Delay,
}
\DeclareAcronym{TEC}{
short=TEC,
long=Total Electron Content,
}
\DeclareAcronym{IPP}{
	short=IPP,
	long=Ionosphere Pierce Point,
}
\DeclareAcronym{NOAA}{
	short=NOAA,
	long=National Oceanic and Atmospheric Administration,
}
\DeclareAcronym{UCR}{
	short=UCR,
	long=University of California-Riverside,
}
\DeclareAcronym{USTEC}{
short=US-TEC,
long=US Total Electron Content,
}
\DeclareAcronym{VNDGNSS}{
	short=VN-DGNSS,
	long=Virtual Network DGNSS,
}
\DeclareAcronym{IOD}{
	short=IOD,
	long=Issue Of Data,
}
\DeclareAcronym{CAS}{
	short=CAS,
	long=Chinese Academy of Sciences,
}
\DeclareAcronym{CNES}{
	short=CNES,
	long=Centre National d'Études Spatiales,
}
\DeclareAcronym{RMS}{
	short=RMS,
	long=Root Mean Square,
}
\DeclareAcronym{SF}{
	short=SF,
	long=Single Frequency,
}
\DeclareAcronym{STD}{
	short=STD,
	long=Standard Deviation,
}
\DeclareAcronym{DF}{
	short=DF,
	long=Dual Frequency,
}
\DeclareAcronym{ICD}{
	short=ICD,
	long=Interface Control Document,
}
\DeclareAcronym{NED}{
	short=NED,
	long={North, East and Down},
}
\DeclareAcronym{WAAS}{
	short=WAAS,
	long=Wide Area Augmentation System,
}
\DeclareAcronym{OPUS}{
	short=OPUS,
	long=Online Positioning User Service,
}
\DeclareAcronym{PDF}{
	short=PDF,
	long=Probability Density Function,
}
\DeclareAcronym{RAPS}{
	short=RAPS,
	long=Risk-Averse Performance-Specified,
}

\DeclareAcronym{VRS}{
	short=VRS,
	long=Virtual Reference Station,
}

\DeclareAcronym{SPP}{
	short=SPP,
	long=Single-frequency Point Positioning,
}

\DeclareAcronym{SPaT}{
	short=SPaT,
	long=Signal Phase and Timing,
}

\DeclareAcronym{TECU}{
	short=TECU,
	long=Total Electron Content Units,
}

\DeclareAcronym{BNC}{
	short=BNC,
	long=BKG NTRIP Client,
}

\DeclareAcronym{ITS}{
	short=ITS,
	long=Intelligent Transportation Systems
}

\DeclareAcronym{USDOT}{
	short=USDOT,
	long=U.S. Department of Transportation
}

\DeclareAcronym{WHU}{
	short=WHU,
	long=Wuhan University
}
\usepackage{fancyhdr}
\fancypagestyle{firstpage}{
	\fancyhf{} % clear all header and footer fields
	\fancyhead[L]{\red PREPRINT: PLEASE CITE THE FINAL VERSION IN 2024 AMERICAN CONTROL CONFERENCE} % Centered header
}

\begin{document}
	\title{Outlier Accommodation for GNSS Precise Point Positioning using Risk-Averse State Estimation}
	\author{Wang~Hu,~Jean-Bernard~Uwineza,
		and~Jay~A.~Farrell% <-this % stops a space
		\thanks{W. Hu (whu027@ucr.edu), J-. B. Uwineza (juwin001@ucr.edu), and J. A. Farrell (farrell@ece.ucr.edu) are with the Dept. of Electrical and Computer Engineering,
			U. of California, Riverside, CA 92521, USA.}
		}
	\maketitle
	\thispagestyle{firstpage}
\begin{abstract}
Reliable and precise absolute positioning is necessary in the realm of  \ac{CAV}. 
\ac{GNSS} provides the foundation for absolute positioning.
Recently enhanced \ac{PPP} technology now offers corrections for \ac{GNSS} on a global scale, 
with the potential to achieve accuracy suitable for  real-time \ac{CAV} applications. 
However, in obstructed sky conditions, \ac{GNSS} signals are often affected by outliers; therefore, addressing outliers is crucial. 
In \ac{GNSS} applications, there are many more measurements available than are required to meet the specification. 
Therefore, measurement selection is important to avoid outlier measurements is of interest. 
The recently developed \ac{RAPS} state estimation optimally selects measurements to minimize outlier risk while meeting a positive semi-definite constraint on performance; at present, the existing solution methods are not suitable for real-time computation and have not been demonstrated using challenging real-world  data or in \ac{RTPPP} applications. 
This article makes contributions in a few directions.
First, it uses a diagonal performance specification, which reduces computational costs relative to the positive semi-definite constraint. 
Second, this article considers \ac{GNSS} \ac{RTPPP} applications.
Third, the experiments use real-world \ac{GNSS} data collected in challenging environments. 
The \ac{RTPPP} experimental results show that among the compared methods: 
all achieve comparable performance in open-sky conditions, 
and all  exceed the \ac{SAE} specification; however, in challenging environments, the diagonal \ac{RAPS} approach shows improvement of 6-19\% over traditional methods. 
Throughout, \ac{RAPS} achieves the lowest estimation risk.

\end{abstract} 	
		
\section{Introduction}\label{sec:intro}

Reliable and accurate real-time positioning is crucial for \ac{CAV} within the framework of intelligent transportation systems \cite{williams2022impact,farrell2023lane}. 
The widespread adoption of \ac{CAV} holds promise to alleviate congestion, decrease both the frequency and severity of accidents, reduce emissions, enhance system resilience by addressing bottlenecks, and offer numerous other benefits \cite{xia2011indirect}. 
Precise and dependable localization and positioning are fundamental requirements for \ac{CAV} applications \cite{williams2022impact,yu2018gps}. Accurate absolute positioning is facilitated by \ac{GNSS}. 
For \ac{CAV} applications, the \ac{SAE} J2945 specification defines the requirement for horizontal 
and vertical position errors to be less than 1.5 m and 3.0 m, respectively, 
with a probability of 68\% \cite{SAEJ2945}. 
However, standalone GNSS receivers, when not supplemented with external corrections, usually achieve positioning accuracy around 10 meters \cite{team2014global}. 
This limitation is caused by \ac{CME} in \ac{GNSS} measurements \cite{farrell2008aided,teunissen2017springer}.
%\ac{CME} are correlated over local areas (i.e., approximately $40\ km$).

Early and traditional methods to counteract the effects of \ac{CME} employed observation space representation corrections. 
This approach is commonly associated with \ac{DGNSS}, where receivers acquire lumped \ac{CME} corrections either from a processing center or nearby base stations. 
 \ac{DGNSS} using pseudorange measurements typically yields an accuracy within the range of 1-3 meters. 
In contrast, \ac{DGNSS} using phase measurements (often referred to as \ac{RTK}) can attain centimeter-level accuracy when integer ambiguities are correctly resolved \cite{teunissen2017springer}. However, these \ac{DGNSS} advantages necessitate extensive local base station infrastructure and do not inherently provide corrections with global coverage.

The \ac{PPP} corrections  have global applicability without the need for users to access local base stations; therefore, they are more appropriate for applications such as \ac{CAV} that require global coverage.
\ac{PPP} uses state space representation, which utilizes models for each individual components of the \ac{CME} \cite{igs2020ssr}. 
These methods were originally developed for post-processing in GPS surveying applications. 
The \ac{IGS} \ac{MGEX} project is a global collaboration,  involving 
\ac{CNES}, \ac{WHU}, \ac{CAS}, \ac{DLR},
and others agencies, to compute and communicate globally relevant \ac{RTPPP} corrections with low latency (approximately 10 seconds)
for  multi-\ac{GNSS} systems (e.g., GPS, GLONASS, Galileo, BeiDou). 
Its products have  recently advanced to the point where they are suitable for real-time \ac{GNSS} operations \cite{felipe2023,igs2020ssr}. 

Each satellite system is designed to provide visibility of 6-12 satellites at any given epoch. 
Each satellite provides measurements on more than one frequency and signal. 
Leveraging multiple \ac{GNSS} systems, frequencies, and signals provides significantly more measurements than the four spatially diverse satellite measurements that are the theoretical requirement for a position (and clock) solution and often more than required to meet the stated accuracy specification. 
When the number of measurements is high, all measurements are outlier-free, and their accuracy is known, utilizing all measurements with optimal state estimation would yield accuracy that far surpasses the requirements of the J2945 specification.

\ac{GNSS} measurements are impacted by multipath and non-line-of-sight effects, as well as spoofing, which can have large-magnitude effects. 
While multipath and non-line-of-sight errors tend to be small under open sky conditions \cite{uwineza2019characterizing}, they become pronounced in challenging environments, such as urban areas (e.g., narrow lanes surrounded by tall buildings and trees). 
In such settings, \ac{GNSS} signals, due to various reflections, are prone to substantial multipath errors. 
Using these outliers measurements within state estimation can distort the state estimate, resulting in over-confidence (i.e., too small covariance) in an incorrect estimate. 
The traditional methods to accommodate outliers in state estimation evaluate measurement residuals using fixed or adaptive thresholds  \cite{fisher1992ical,frank1997survey,patton1994review}. 
In other applications, such as least squares, additional outlier rejection methods have been explored, including least soft-threshold squares \cite{wang2015robust} and median least squares \cite{rousseeuw1984least}. 
These approaches primarily aim to detect and disregard outliers or to de-weight measurements based on the magnitude of their residuals.
These approaches may far surpass the specification requirements when many outlier-free measurements are available and may not meet the performance specification in circumstances where many measurements are de-weighted or removed.
They do not have a mechanism to balance between minimizing measurement risks and satisfying performance specifications.

An alternative optimization-based approach that selects a subset of measurements to achieve a performance specification constraint while minimizing risk is introduced as \ac{RAPS} estimation in \cite{Aghapour_TCST_2019,aghapour2018outlier,rahman2018outlier}. 
The \ac{RAPS} optimization problem under study necessitates that the performance specification constraint be positive semi-definite. 
An alternative method, employing a diagonal performance specification, has been mentioned but has not yet been evaluated. 
A diagonal performance specification eliminates the computational overhead associated with verifying the positive definiteness of a matrix, which has a time complexity of O($n^3$)  in every step of searching for the measurement selection vector in each epoch. 
In addition, prior research showcased \ac{RAPS} within the context of \ac{DGNSS} with artificially imposed outliers. 
\ac{RAPS} has not yet been evaluated in cutting-edge \ac{RTPPP} applications or using real-world \ac{GNSS} data acquired in challenging environments.

Many legacy and low-cost receivers that are currently typical in automotive applications typically only support single frequency. 
Single frequency \ac{GNSS} measurement types include pseudorange, carrier phase, and Doppler. 
The advantages of utilizing phase measurements hinge on the accurate mitigation of \ac{CME} and correct estimation of integer ambiguities. 
When outliers are present, \ac{GNSS} carrier phase measurements become susceptible to outages and cycle slips that complicate ambiguity estimation. 
Furthermore, phase measurement-based PPP-RTK demands enhanced corrections for atmospheric delays, necessitating slant \ac{TEC} and tropospheric messages. 
These are expected in the third stage scheduled by \ac{IGS}, which is currently only in its second stage \cite{igs2020ssr}. 

This paper's contribution lies in demonstrating outlier accommodation for multi-\ac{GNSS} single frequency \ac{RTPPP}, focusing on pseudorange and Doppler, using a diagonal performance-specified \ac{RAPS} approach.
The experiments use real-world data. 
This paper is structured as follows.
Section \ref{sec:model} reviews the \ac{GNSS} observation models for pseudorange and Doppler. 
Section \ref{sec:ppp} discusses the \ac{RTPPP} sources and models for mitigating the \ac{CME}. 
Section \ref{sec:raps} compares the \ac{RAPS}  and  \ac{TD} approaches in state estimation. 
Section \ref{sec:exp} presents and analyzes the \ac{RAPS} performance based on a moving platform experiment. 
Section \ref{sec:conclu} concludes the paper and outlines potential directions for future research.

\section{GNSS Observation Model}\label{sec:model}
State estimation will be performed using \ac{RTPPP} corrected multi-\ac{GNSS} pseudorange and Doppler measurements. 
This article focuses on the GNSS constellations that use code division multiple access: GPS, Galileo, and BeiDou.

%\ac{GNSS} is comprised of various constellations using \ac{CDMA}: \ac{GPS}
For each \ac{GNSS} constellation $\gamma$, the single frequency pseudorange observation of satellite $s$ at frequency $j$ tracked by receiver $r$ at time $t_r$ is modeled as \cite{hu2022using, li2015accuracy}
\begin{align}
	\begin{split}
	\rho^s_{r,j} (t_r) &= R(\Boldp_r, \hat{\Boldp}^s)+ dt_r^{\gamma} + M_r^s + \eta^s_{\rho} \\
	&~~~~- (dt^s - b^s) + T^s_r + I^s_{r,j} + E_r^s
	\end{split} \label{eqn:code_model}
\end{align}
where $R(\Boldp_r, \hat{\Boldp}^s) = \| \Boldp_r-\hat{\Boldp}^s\|$ is the geometric range between satellite and receiver, 
$\hat{\Boldp}^s \in \Re^3$ is the computed satellite position, and 
$\Boldp_r \in \Re^3$ is the receiver position; 
$dt_r^{\gamma}$ is the receiver clock offset for constellation $\gamma$ in meters; 
$dt^s$ is the satellite clock error in meters; 
$b^s$ is the satellite hardware biases in meters;
$T^s_r$ is the tropospheric delay; 
$I^s_{r,j}$ is the ionospheric delay at $j$ frequency; 
$E_r^s$ is the ephemeris error in the computed satellite position $\hat{\Boldp}^s$;
$M_r^s$ is the multipath error; 
$\eta^s_{\rho}  \sim \mathcal{N}(0,(\sigma^s_{\rho})^2)$ is the pseudorange measurement noise.
The hardware biases are normally stable over time \cite{sardon1994estimation}.
\ac{RTPPP} techniques significantly reduce the magnitude of the \ac{CME} terms.
These are listed in the second line of eqn. \eqref{eqn:code_model}.

After compensating for the satellite velocity and clock drift, which can be computed using ephemeris, the corrected Doppler measurement can be modeled as \cite{rahman2018ecef2, farrell2008aided}
\begin{align}
	\delta D^s_r &= (\1^s_{r})^\top \, \Boldv_r + \dot{dt}_r + \epsilon^s_D
\end{align}
where $\Boldv_r \in \mathbb{R}^3$ is the receiver velocity in meter per second, 
$\dot{dt}_r$ is the receiver clock drift in meter per second,
$\1^s_{r}$ is the line-of-sight vector from the satellite to the receiver which is defined as
\begin{align}
	\1^s_{r} &= \frac{\Boldp_r-\Boldp^s}{\| \Boldp_r-\Boldp^s \|},
\end{align}
and $\epsilon^s_{r,D} \sim \mathcal{N}(0,(\sigma^s_{D})^2)$ is the Doppler measurement noise.

\section{Real-time PPP Corrections}\label{sec:ppp}

In eqn. \eqref{eqn:code_model},  without using external corrections,  the satellite position and clock error are computed using the broadcast ephemeris, the satellite hardware bias is roughly corrected by timing group delay from the ephemeris, and troposphere and ionosphere are compensated by standard models. 
This approach yields a range measurement accuracy of about $10$ m.
\ac{RTPPP} corrections for multi-\ac{GNSS} have been developed and maintained by \ac{IGS} MGEX with the goal of achieving enhanced accuracy.

This section discusses the choice of \ac{RTPPP} corrections used herein to mitigate the \ac{CME} effects. 
The PPP corrections are communicated via parameters for models of each component of the \ac{CME}. 
They demonstrate efficacy on a global scale.
Various sources are available. 
These sources have been comprehensively defined and examined in \cite{hu2022using, igs2020ssr, li2012new, sch2016sinex}. 
The real-time orbit and clock products are corrections to the satellite position and clock error from broadcast ephemeris. 
In this study, we choose to use those from WHU. 
Satellite hardware bias could be corrected using either differential code bias or \ac{OSB} format corrections.
The multi-\ac{GNSS} \ac{OSB} products are more convenient.  
In this study, the \ac{OSB} products are provided by CAS.
Tropospheric delay is corrected by the global {\em IGGtrop} empirical model.
Ionospheric delay is compensated using the real-time global vertical \ac{TEC} product provided by CNES.
The model and the reasons for choosing it are discussed in Section IV.D of \cite{hu2022using}.
Table \ref{tab:ppp_models} summarizes the choices.

The \ac{RTPPP} corrections related to satellite orbit and clock, satellite hardware bias, and tropospheric delay can achieve centimeter-level accuracy. 
The assessment of the CNES vertical \ac{TEC} product indicates that the \ac{RMS} ranges from $2.07$ to $6.15$ \ac{TEC} units, 
which, when scaled using the GPS L1 frequency, corresponds to $0.34$ to $1.00$ meters \cite{nie2019quality}.

After mitigating the \ac{CME}s, the corrected pseudorange model can be represented as:
\begin{align}
	\label{eq:corrected_rho}
	\delta \rho^s_r (t_r) &= R(\Boldp_r, \hat{\Boldp}_c^s)+dt_r^{\gamma} + M_r^s + \epsilon^s_r
\end{align}
where $\hat{\Boldp}_c^s$ denotes the computed satellite position using the PPP orbit products
and 
$\epsilon^s_r = \Delta^s_r + \eta^s_{\rho}$ where $\Delta^s_r$ represents the residual of \ac{CME}s.

\begin{table}[tb]
	\centering
	\caption{Correction sources and models for multi-\ac{GNSS} single frequency \ac{RTPPP}}
	\begin{tabular}{c|c|c}
		\hline
		\ac{CME} \T					& \ac{CME} Symbol							& Correction sources or models \\ \hline
		Satellite orbit \T 		& \multirow{2}{*}{$\Boldp^s$ and $dt^s$} 	& \multirow{2}{*}{ WHU products \cite{guo2016precise}} \\
		and clock       		&   & \\ \hline
		Satellite    \T  		& \multirow{2}{*}{$b^s$}   					& \multirow{2}{*}{CAS GIPP \ac{OSB} product \cite{sch2016sinex,wang2020gps}} \\
		hardware bias   &      &    \\ \hline
		Tropospheric  \T 		& \multirow{2}{*}{$T^s_r$}    				& \multirow{2}{*}{ IGGtrop \cite{li2012new}}  \\
		delay    &       & \\ \hline
		Ionospheric  \T  		& \multirow{2}{*}{$I^s_{r,j}$}     			& \multirow{2}{*}{CNES vertical \ac{TEC} product \cite{roma2016real}}                \\
		delay           &      &  \\ \hline
	\end{tabular}
	\label{tab:ppp_models}
\end{table}

During the state estimation process, the measurement noise and multipath combined is assumed to be white and normal, $( M_r^s +\epsilon^s_r) \sim \mathcal{N}(0,(\sigma^s_{\rho})^2 ) $. 
This is reasonable in open sky conditions where multipath effects are generally small \cite{khanafseh2018gnss,uwineza2019characterizing}.
However, in urban and residential areas, multipath errors can unexpectedly surge to tens of meters. 
As there's no established model for multipath errors, when they are large, they are treated as outliers. In this study, the variance of pseudorange is computed by $(\sigma^s_{\rho})^2 = \sigma^2_b+(M^s_a\,\sigma_a)^2+\sigma^s_{\rho}$ where $\sigma^2_b$ is the variance of \ac{PPP} corrections excluding ionosphere correction, $\sigma^2_a$ is the variance of vertical \ac{TEC} units of the ionosphere product, and $M^s_a$ is the mapping factor (see Sec. 4.5 in \cite{igs2020ssr}) for satellite $s$. The standard deviation values for the experiment are summarized in Table \ref{tab:var}.

\begin{table}[tb]
	\centering
	\caption{Standard deviation values for measurement noise model.}
	\begin{tabular}{c|c|c|c}
		\hline
		$\sigma_b$\T & $\sigma_a$    & $\sigma^s_{\rho}$   & $\sigma^s_{D}$ \\ \hline
		0.1 m\T & 6.15 TEC & 0.9 m & 1.414 m/s  \\ \hline
	\end{tabular}
	\label{tab:var}
\end{table}

\black

\section{ State  Estimation}\label{sec:raps}
This section describes the state vector and system model.

The state vector is defined as
\begin{equation}
	\Boldx = [\Boldp_r,\,\Boldv_r,\,\Bolda_r,\,\Bolddt_r,\,\dot{dt}_r]^\top \in \Re^n
\end{equation}
where $\Bolda_r \in \Re^3$ denotes the receiver acceleration,  
$\Bolddt_r \in \Re^{n_s}$ represents the vector of receiver clock offsets  for the  $n_s$ different constellations, and $\dot{dt}_r$ is the scalar rate of clock offset drift.
The state at time $t_k=k\,T$ will be denoted as $\Boldx_k$ where $T$ is the sampling interval.

The time update portion of the discrete-time model 
\begin{align}
	\Boldx_{k+1} = \BoldPhi\,\Boldx_k + \Boldw_k
\end{align}
is a standard position, velocity, acceleration approach
where the state transition matrix is
\begin{align}
	\BoldPhi = \begin{bmatrix}
		\BoldI_{3 } & T\,\BoldI_{3 } & \frac{1}{2}T^2\,\BoldI_{3 } & \0 & \0 \\
	\0	& \BoldI_{3 } & T\,\BoldI_{3 } & \0 & \0\\
	\0	& \0 & \BoldI_{3} & \0 & \0\\
	\0	& \0 & \0 & \BoldI_{n_s} &  \BoldT_{n_s} \\
	\0	& \0 & \0 & \0 & 1
	\end{bmatrix}
\end{align}
and $\Boldw_k \sim \mathcal{N}(\0,\BoldQ)$ is the white Gaussian process noise where the discrete-time process noise covariance matrix $\BoldQ$ can be computed from the continuous-time power spectral density using eqn. (4.110) in \cite{farrell2008aided}, $\BoldI_{q}$ represents the identity matrix of dimension $q$, 
and $\BoldT_{n_s}$ is a column vector with $n_s$ elements all having value of $T$. 
The symbol $\0$ is a conformal matrix containing all zeros.

The measurement vector at time $k$ is:
\begin{align}
	\Boldy_k =  \Boldh(\Boldx_k) + \Boldeps_k
\end{align}
where $\Boldy_k = [\delta \rho_r^1,\, ...,\,\delta \rho_r^m, \delta D_r^1,\,...,\,\delta D_r^m]^\top\in\Re^{2\,m}$ 
contains corrected pseudorange and Doppler measurements for $m$ satellites, 
$\Boldh(x_k) = [h_{\rho}^1(x_k),\,...,\,h_{\rho}^m(x_k),\,h_D^1(x_k),\,...,\, h_D^m(x_k)]^\top$ represents the measurement models where
\begin{align}
	h_{\rho}^m(\Boldx_k) &= R(\Boldp_r, \Boldp_c^s)+c\,dt_r^{\gamma}~\text{and} \nonumber \\
	h_D^m(\Boldx_k) &= (\1^s_{r})^\top \cdot \Boldv_r + c\,\dot{dt}_r, \nonumber
\end{align}
and $\Boldeps_k \sim \mathcal{N}(\0,\BoldR_k)$ represents the Gaussian measurement noise. The pseudorange and Doppler measurement noises are mutually uncorrected, each is white, and both are uncorrelated between satellites; therefore,
the covariance matrix $\BoldR_k = diag\,( [(\sigma^1_{\rho})^2,\,...\,(\sigma^m_{\rho})^2,\,(\sigma^1_D)^2,\,...\,(\sigma^m_D)^2])$ is diagonal and invertible.

The time propagation model for the state estimate is:
\begin{align}
	\hat{\Boldx}_{k+1}^- &= \BoldPhi\,\hat{\Boldx}_k^+, \\
	\BoldP^-_{k+1} &= \BoldPhi\,\BoldP_k^+ \, \BoldPhi^\top + \BoldQ
\end{align}
where $\hat{\Boldx}_k^-$ and $\hat{\Boldx}_k^+$ denote the prior and posterior state estimate, 
and $\BoldP_k^-$ and $\BoldP_k^+$ denote the prior and posterior state estimate error covariance matrix. 
The \ac{PDF} of prior state $\Boldx_k$ is assumed to be Gaussian $\Boldx_k \sim \mathcal{N}(\hat{\Boldx}^-_k,\BoldP^-_k)$. 
Outliers are not directly involved in the time propagation process since they only affect the measurements, the outlier accommodation is considered in the measurement update process.

\section{Measurement Outlier Accommodation}
This section reviews the traditional \ac{TD} outlier rejection approach and presents  \ac{RAPS} estimation \cite{aghapour2018outlier} modified for the diagonal performance specification constraint.

Measurement selection will be implemented by use of a vector $\Boldb = [b_1,\,...,\,b_{2\,m}]^\top$ with $b_i \in \{0,\,1\}$, see \cite{carlone2014selecting}. 
The vector $\Boldb$ is used to select a certain measurement to be used ($b_i = 1$) in the measurement update or to be ignored ($b_i = 0$). 
The maximum-a-posteriori state estimation, for a fixed value of $\Boldb$, yields the nonlinear least squares problem
\begin{align}\label{eqn:state_estimate}
	\hat{\Boldx}_k^+ &= \underset{\Boldx_k}{\text{argmin}}~C(\Boldx_k,\,\Boldb)
\end{align}
where the cost function (see \cite{aghapour2018outlier})
\begin{align}
	C(\Boldx_k,\,\Boldb) &= \left\|\BoldSigma_{\BoldP} (\Boldx_k - \hat{\Boldx}_k^-) \right\|^2 + \left\|\BoldSigma_{\BoldR} \, \BoldPsi(\Boldb)\left(\Boldh(\Boldx_k) - \Boldy_k\right) \right\|^2 \nonumber \\
&= \left\|\BoldSigma_{\BoldP} (\Boldx_k - \hat{\Boldx}_k^-) \right\|^2 + \sum_{i=1}^{2\,m} \frac{b^2_i}{\sigma^2_i} \Bolde_i\,\Bolde_i^\top \left(\Boldh(\Boldx_k) - \Boldy_k\right)^2 \nonumber\\
&= \left\|\BoldSigma_{\BoldP} (\Boldx_k - \hat{\Boldx}_k^-) \right\|^2 + \sum_{i=1}^{2\,m} \frac{b_i}{\sigma^2_i} \Bolde_i\,\Bolde_i^\top \left(\Boldh(\Boldx_k) - \Boldy_k\right)^2 \nonumber
\end{align}
where $b_i^2=b_i$, $\BoldPsi(\Boldb) = diag(\Boldb)$, 
${\BoldSigma_{\BoldR}}^\top \, \BoldSigma_{\BoldR} = \BoldR_k^{-1} $, ${\BoldSigma_{\BoldP}}^\top \, \BoldSigma_{\BoldP} = (\BoldP_k^-)^{-1} $,
$\sigma^2_i$ is the $i$-th diagonal element of $\BoldR_k$, and $\Bolde_i$ is the $i$-th standard basis vector. For any fixed value of $\Boldb$, the minimum of $C(\Boldx_k,\,\Boldb)$ as a function of $\Boldx_k$ quantifies the risk associated with the measurements selected by  $\Boldb$.

The posterior information matrix, when using the measurements selected by 
$\Boldb$, is:
\begin{align}
	\BoldJ_{\Boldb}^+ &= \BoldH^\top \BoldPsi(\Boldb)^\top \BoldR^{-1}\BoldPsi(\Boldb)\BoldH
	+  \BoldJ_k^- \nonumber \\
	&= \sum_{i=1}^{2\,m} \frac{b_i}{\sigma^2_i} \Boldh^\top_i \Boldh_i + \BoldJ_k^-  \label{eqn:post_info}
\end{align}
where $\Boldh_i$ is the $i$-th row of $\BoldH = \frac{\partial}{\partial\,\Boldx}\Boldh(\Boldx)|_{\Boldx = \Boldx_k} \in \mathbb{R}^{2\,m\times n}$ and $\BoldJ_k^- = (\BoldP_k^-)^{-1}$ is the prior information matrix. 
The corresponding posterior covariance matrix is $\BoldP^+_k = (\BoldJ^+_{\Boldb})^{-1}$.

Two strategies are considered for determining $\Boldb$.

\subsection{Threshold Decisions}
The traditional approach for detecting and removing outliers in state estimation employs a threshold test on each scalar  measurement residual. 
The vector of residuals is 
 computed as 
 \begin{align}
     \Boldr_k =  \Boldy_k - \Boldh(\hat{\Boldx}_k^-).
 \end{align}
 Let $r_i$ denote the $i$-th element of $\Boldr_k$.
The selection of $\Boldb$ is determined by
\begin{equation} \label{eqn:threshold_test}
	b_{i} = \left\{
	\begin{array}{ll}
		0, & \mbox{ when }{|r_i|} \ge \lambda\sigma_{r_i}\\
		1, & \mbox{ when }{|r_i|} <   \lambda\sigma_{r_i}
	\end{array}
	\right.
\end{equation}
where $\lambda>0$ is the decision threshold and
$\sigma_{r_i}^2= {\Boldh_i \, \BoldP^-_k \, \Boldh_i^\top + \sigma_{i}^2}$ 
is the covariance of  $r_i$ (see e.g., \cite{fisher1992ical,frank1997survey,patton1994review}). 
For the selection vector $\Boldb$ that results, the optimal estimate $\hat{\Boldx}_k^+$ and its posterior information $\BoldJ_{\Boldb}^{+}$ are  
the solutions of eqn. \eqref{eqn:state_estimate} and eqn. \eqref{eqn:post_info}.

The threshold test approach requires the designer to select a decision threshold  $\lambda$. 
Regardless of the existence of outliers, $\BoldJ_{\Boldb}^+$ is maximized by selecting all measurements, which would be fine if no outliers existed. 
When outliers exist, the state estimate and error covariance matrix are only correct when all outliers are removed at each epoch. 
However, in practice, no detection method is flawless. 
Missed outlier detection corrupts the state estimate mean $\hat{\Boldx}_k^+$ 
and causes the covariance matrix to underestimate the true uncertainty.
This leads to an inconsistency between the mean and variance, rendering subsequent outlier decisions unreliable.
Moreover, the \ac{TD} approach has a fixed decision criteria that makes decisions without regard to the performance specification. 

\subsection{\ac{RAPS} Approach}
Instead of focusing on detecting outliers, \ac{RAPS} seeks to select a subset of measurements to use that minimize the risk, as quantified by $C(\Boldx_k,\,\Boldb)$, while achieving the defined accuracy specification. 
This accuracy specification is realized by imposing a constraint on the posterior information matrix within an optimization framework. 
The previous research in \cite{Aghapour_TCST_2019,aghapour2018outlier}  employed a semi-definite matrix constraint (i.e., $\BoldJ_{b}^+ \ge  \BoldPsi(\BoldJ_d)$). 
Herein, we consider an alternative diagonal performance constraint.
Use of the diagonal constraint reduces the computational cost of the optimization.
This formulation has not been evaluated in the literature previously.
Moreover, earlier studies mainly analyzed the impact on positioning performance as a function of outlier magnitudes by artificially imposing outliers. 
The efficacy of \ac{RAPS} is particularly intriguing in the context of real-world \ac{RTPPP} applications.

The \ac{RAPS} optimization problem, utilizing the diagonal performance specification, is expressed as:
\begin{equation} \label{eqn:RAPS_Problem1}
	\left.
	\begin{aligned}
		\Boldx_k^+, \Boldb^\star  = \,&\underset{\Boldx_k,\Boldb}{\text{argmin}} ~ C(\Boldx_k,\Boldb)  \\ 
		&\text{s.t.:} \ \ \mbox{diag}\,(\BoldJ_{b}^+) \ge \BoldJ_d \\
		&\ \ \ \ \ b_i \in \{0,\,1\} \ \text{for} \ i = 1,\,...,\,2\,m
	\end{aligned}
	\right\}
\end{equation}
where $\BoldJ_d$ represents a positive vector defining the performance constraint for each element of the state vector. 
The user-defined performance constraint parameter $\BoldJ_d$ to achieve the \ac{SAE} J2945 is discussed in  Section III-B of \cite{Aghapour_TCST_2019}. 
The parameters to achieve this specification are 
defined by
$$\BoldJ_p = [1.929,\,1.929,\,0.121]^\top \mbox{ and } \BoldJ_v = [1.389,\,1.389,\,0.347]^\top$$
with $\BoldJ_d = [\BoldJ_p^\top, \BoldJ_v^\top]^\top$.
The first two elements of $\BoldJ_p$ and $\BoldJ_v$ specify the  local tangent plane north and east direction information bounds, while the third element applies to the vertical direction.

%\red Solution of ... is discussed in....the dissertation \black

Starting from eqn. \eqref{eqn:post_info}, the constraint $\mbox{diag}(\BoldJ_{b}^+) \ge \BoldJ_d$ can be expressed as
\begin{equation} \label{eqn:diag_spec}
	\sum_{2\,m}^{i=1}\frac{b_i}{\sigma^2_i}\ \text{diag}\,(\Boldh_i^\top\,\Boldh_i) +\,\text{diag}\,(\BoldJ_k^-) \ge \BoldJ_d.
\end{equation}

The solution to the \ac{RAPS} optimization problem for the  matrix constraint $\BoldJ_{b}^+ \ge \BoldPsi(\BoldJ_d)$ is discussed in \cite{Aghapour_TCST_2019}.
This case requires a positive definiteness check on $(\BoldJ_{b}^+ -  \BoldPsi(\BoldJ_d))$,  which is an  $O(n^3)$ computation. 
The solution in  \cite{Aghapour_TCST_2019} employed an exhaustive search strategy for the binary vector $\Boldb$, which has comptational cost $O(m!)$. In contrast, the diagonal case in eqn. \eqref{eqn:diag_spec} only requires $n$ scalar inequality evaluations.
An efficient solution to Problem \eqref{eqn:RAPS_Problem1}, using a block coordinate method with polynomial time-complexity, is described in Sec. 6 of \cite{hu2024optimization} and is utilized in this paper.

\section{Experiment Performance and Analysis}\label{sec:exp}

This section presents experimental results for the \ac{GNSS} positioning using \ac{RAPS} and \ac{RTPPP} corrections. 
Results from \ac{TD}, \ac{RAPS}, and the standard \ac{EKF} are compared. 
To be comparable with automotive applications, a low-cost u-blox F9P receiver is used to acquire the raw \ac{GNSS} measurements on a moving platform.
The analysis of multi-GNSS \ac{RTPPP} is performed using GPS L1, Galileo E1, and Beidou B1 signals. 
The elevation cut-off is $10$ degrees. 
The experimental vehicle navigated through urban and residential zones near University of California, Riverside, 
traversing areas characterized by narrow lanes with tall trees lining the road. 
Throughout the experiment, the u-blox maintained communication with a nearby base station. 
The $ground \ truth$ trajectory was established using u-blox's native multi-GNSS, dual-frequency, integer-fixed \ac{RTK} solution to achieve centimeter accuracy. \footnote{The experiment code is public on GitHub: \url{github.com/Azurehappen/Outlier_Accommodation_GNSS_ACC2024}}

\subsection{Estimators}
The results are compared for three estimators.
The time propagation for all estimators is identical. 
The measurement update for all estimators determines the posterior state $\Boldx_k^+$ by minimizing the cost function in eqn. \eqref{eqn:state_estimate}, but choose $\Boldb$ differently.
\begin{description}
\item[EKF:] Uses all measurements (i.e., $\Boldb = \mathbf{1}$).
Since it uses all measurements both the cost and the posterior information matrix will be maximized over all choices of $\Boldb$.

\item[TD:] Defines $\Boldb$ using eqn. \eqref{eqn:threshold_test} with 
  $\lambda = 1.5$.
  
\item[RAPS:] Chooses $\Boldb$ during the solution of the optimization problem to satisfy the performance specification constraint, as stated in eqn. \eqref{eqn:RAPS_Problem1}.
\end{description}

\subsection{Metrics}
The analysis of positioning performance between different estimators will be based on the following metrics:
\begin{itemize}
	\item North, East and Down frame coordinate errors: $\delta p_N$, $\delta p_E$, $\delta p_D$;
	\item Horizontal positioning error: $HE = \sqrt{\delta p^2_N + \delta p^2_E}$;
	\item Vertical positioning error: $VE = |\delta p_D|$; and
    \item Estimation risk as quantified by $C(\Boldx^+_k,\,\Boldb)$.
\end{itemize}

\subsection{Results}
\begin{figure}[bt]
	\centering
	\includegraphics[width=\linewidth]{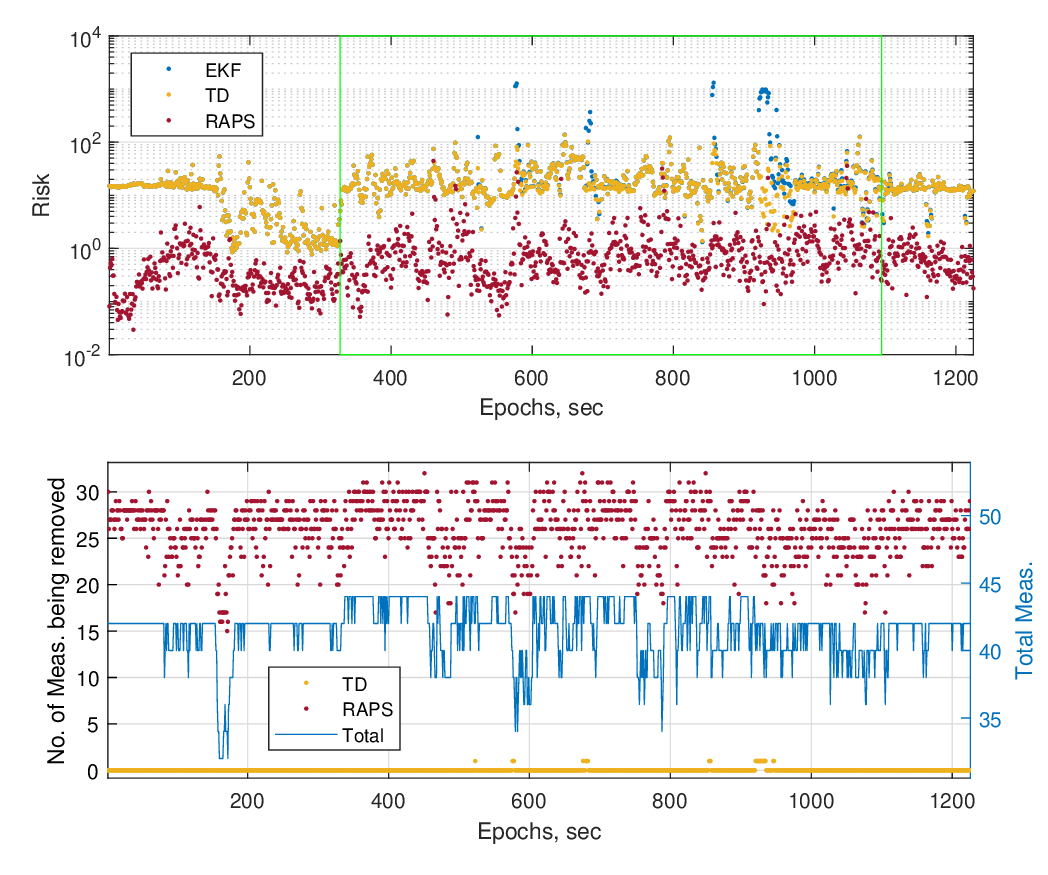}
	\caption{The top graphs display experimental results for the risk from \ac{EKF} in blue, \ac{EKF} with \ac{TD} in orange, and \ac{RAPS} in red. 
 The solid green box indicates the period during which the driving was under non-clear view of sky conditions. 
 The bottom graphs display the number of measurements removed by either \ac{RAPS} or \ac{TD} (left $y$-axis) and the total number of measurements available (right $y$-axis). 
 The number of measurements available is equivalent to the number used by the \ac{EKF}. }
	\label{fig:cost}
\end{figure}

The performance of the three estimators is evaluated based on their estimation risk and positioning accuracy. 
The top graphs in Fig. \ref{fig:cost} display the risks across all epochs.
The time period highlighted by the green box represents the vehicle's journey through the residential areas, in which the narrow lanes with tall trees increase the probability of encountering frequent and significant outlier measurements.  
This time period is referred to as the ``obstructed view'' condition. 
In contrast, other epochs offer a clear view of the open sky.

The red graph indicates that \ac{RAPS} consistently achieves the lowest risk. 
Both \ac{EKF} and \ac{TD} exhibit risks that are significantly higher than that of \ac{RAPS}. 
In many epochs, the risk difference between \ac{EKF} and \ac{TD} is small, so the EKF data is covered by the TD data.
This shows that for most time epochs, most residuals passed the TD test, even with the reasonable small value of $\lambda,$
regardless of whether the conditions were open sky or not. 
Notably, during specific epochs (around 600s and 1000s), \ac{EKF} demonstrates a pronounced risk, 
whereas \ac{TD} exhibits a reduced risk due to its ability to reject severe outliers.

The bottom graphs in Fig. \ref{fig:cost} display the number of measurements discarded by  \ac{RAPS} or \ac{TD}. \ac{RAPS} selects a subset of measurements to minimize risk while achieving the performance specifications. 
The solution is feasible for all epochs, even while removing many more measurements than \ac{TD}. 
At each epoch, 
RAPS adapts both the number of satellites removed and the specific satellites removed based on the number and geometry of the available measurements as well as their specific residuals. 
This selection results in the reduced risk demonstrated in the top figure while achieving the performance specification at all times.

\begin{figure}[bt]
	\centering
	\includegraphics[width=\linewidth]{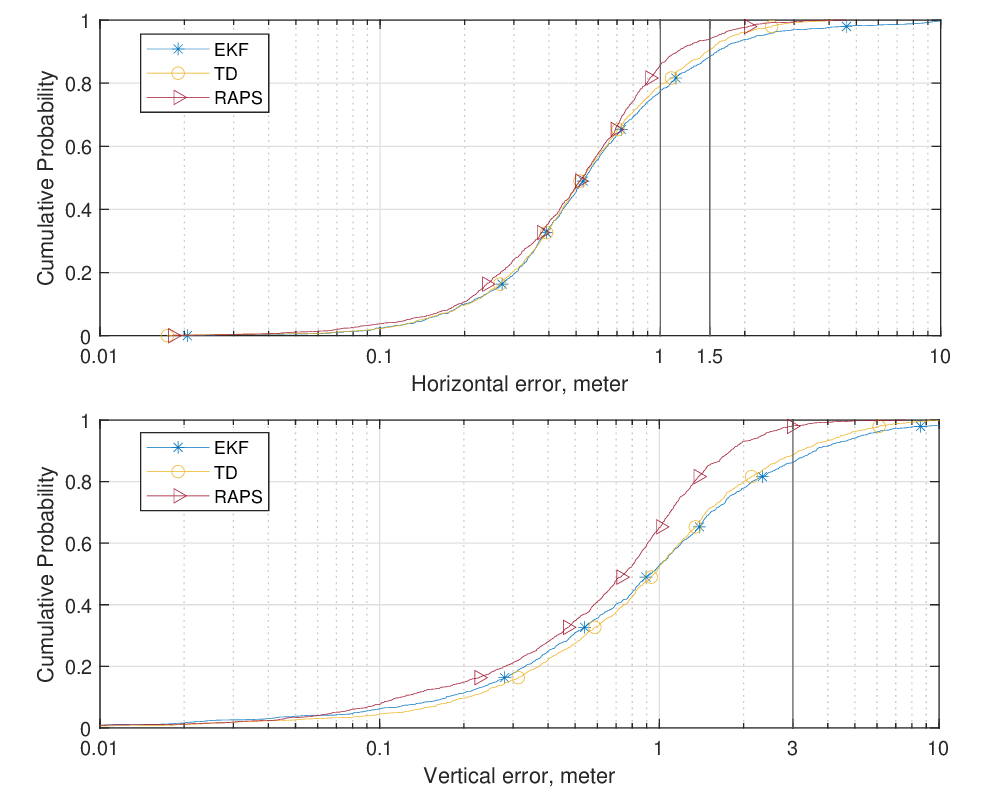}
	\caption{The cumulative error distribution of horizontal (top figure) and vertical (bottom figure) positioning errors for \ac{EKF} in blue, \ac{EKF} with \ac{TD} in orange, and \ac{RAPS} in red for the entire experiment trajectory.}
	\label{fig:cdf}
\end{figure}

Fig. \ref{fig:cdf} presents the cumulative probability distribution for both horizontal and vertical positioning errors, 
using a consistent color scheme with Fig. \ref{fig:cost}.
Table \ref{fig:status_rms} and \ref{fig:status_pr} provide a summary of the positioning performance statistics from the different estimators under various conditions. 
Table \ref{fig:status_rms}  includes the mean, \ac{RMS}, and maximum values of the norm and absolute value of the horizontal and vertical errors.
Table \ref{fig:status_rms} displays the probability of the positioning errors being less than the specified threshold corresponding to the SAE specification. 
The best results are highlighted in bold. 
Throughout the entire experimental period, \ac{RAPS} delivers superior results, 
with the exception of the maximum horizontal error, 
which exceeds that of \ac{TD} by 0.39 m. \ac{RAPS} showcases an advantage of approximately 3.51-12\% across all probability metrics. 
It is notable that all maximum errors originate from the non-clear sky condition. 
Segmenting the comparison between non-clear sky and open sky conditions,
single frequency \ac{RTPPP} performance of these three estimators all surpass the SAE requirements under both conditions. 
Further analysis is as follows:
\begin{itemize}
	\item Obstructed View: \ac{RAPS} enhances the horizontal accuracy at both 1-meter and 1.5-meter levels by approximately 6-12\% compared to \ac{TD}. 
	While \ac{TD} outperforms \ac{EKF}, the improvements are relatively smaller. 
	\ac{TD} struggles to eliminate sufficient outliers for better performance. 
	Its inability to detect certain outliers results in close risks comparable to those of \ac{EKF}, as depicted in Fig. \ref{fig:cost}. 
	In this sky condition, \ac{RAPS} achieves the lowest mean and \ac{RMS}. 
	Both \ac{TD} and \ac{RAPS} exhibit fewer and smaller maximum positioning errors in challenging scenarios compared to \ac{EKF}.
	
	\item Open sky:
	The performance disparities among the three estimators are less pronounced under open sky conditions, which typically present minimal outliers. 
    If it were known that there were no outliers and the goal was to minimize state estimation error, then using all measurements is the correct approach, which is what the EKF is doing. 
    When the goal is to achieve a state specification, then using all measurements invokes extra risk. 
    RAPS will only use the best set of measurements to achieve the specification at the lowest risk.
    Therefore, \ac{RAPS} is not designed to lead in statistical performance under open sky (no outlier) conditions, yet it does comparably well by minimizing risk subject to its performance constraint.
    Because both \ac{EKF} and \ac{TD} use essentially all measurements,  their posterior state and error covariance remain similar to each other.
    Also, both their risk and (theoretical) posterior information will be higher than that of \ac{RAPS}.
\end{itemize}

\begin{table}[bt]
	\centering
	\begin{tabular}{cc|ccc}
		\hline
		\multicolumn{2}{c|}{\multirow{2}{*}{}}  						& Mean (m)\T 					& RMS (m)\T 					& Max (m)\T \\
		\multicolumn{2}{c|}{}                   						& HE / VE \T  					& HE / VE\T 					& HE / VE\T \\ \hline
		\multicolumn{1}{c|}{\multirow{3}{*}{Overall}}    	& EKF\T 	& 0.87 / 1.58 					& 1.52 / 2.617 					&  12.39 / 20.19 \\
		\multicolumn{1}{c|}{}     							& TD\T 		& 0.70 / 1.39 					& 0.91 / 2.03 					& \textbf{4.12} / 10.54 \\
		\multicolumn{1}{c|}{}  							& RAPS\T 	& \textbf{0.63} / \textbf{0.90} & \textbf{0.80} / \textbf{1.19} & 4.61 / \textbf{7.57} \\ \hline
		\multicolumn{1}{c|}{\multirow{3}{*}
		{\begin{tabular}[c]{@{}c@{}}Obstructed\\ View\end{tabular}}} 	
															& EKF\T 	& 1.14 / 2.09  					& 1.88 / 3.29  					& 12.39 / 20.19 \\
		\multicolumn{1}{c|}{}    							& TD\T  	&  0.88 / 1.75  				& 1.08 / 2.46  					& \textbf{4.12} / 10.54 \\
		\multicolumn{1}{c|}{}       						& RAPS\T 	& \textbf{0.73} / \textbf{0.95} & \textbf{0.90} / \textbf{1.28} & 4.61 / \textbf{7.57} \\ \hline
		\multicolumn{1}{c|}{\multirow{3}{*}{Open sky}}    	& EKF\T 	& \textbf{0.41} / \textbf{0.75} & \textbf{0.49} / \textbf{0.96} & 2.59 / \textbf{4.81} \\
		\multicolumn{1}{c|}{}     							& TD\T 		& \textbf{0.41} / 0.78  		& 0.50 / 0.98  					& 2.59 / \textbf{4.81} \\
		\multicolumn{1}{c|}{}     							& RAPS\T 	&   0.48 / 0.80  				&  0.58 / 1.03 					& \textbf{2.13} / 4.89  \\ \hline
	\end{tabular}
	\caption{Horizontal and vertical position error magnitudes.}
	\label{fig:status_rms}
\end{table}

\begin{table}[bt]
	\centering
	\begin{tabular}{cc|ccc}
		\hline
		\multicolumn{2}{c|}{\multirow{2}{*}{}}    						& Prob. of\T  							& Prob. of\T   							& Prob. of\T    \\
		\multicolumn{2}{c|}{}                     						& \multicolumn{1}{l}{HE $\leq$ 1.0 m} 	& \multicolumn{1}{l}{HE $\leq$ 1.5 m} 	& \multicolumn{1}{l}{VE $\leq$ 3.0 m} \\ \hline
		\multicolumn{1}{c|}{\multirow{3}{*}{Overall}}   & EKF\T 		& 77.55\% 								& 88.24\% 								& 86.12\% \\
		\multicolumn{1}{c|}{}     						& TD\T  		& 79.76\%  								& 90.86\% 								& 88.65\% \\
		\multicolumn{1}{c|}{} 							& RAPS\T 		&  \textbf{86.04\%}  					& \textbf{93.96\%} 						& \textbf{98.12\%} \\ \hline
		\multicolumn{1}{c|}{\multirow{3}{*}
		{\begin{tabular}[c]{@{}c@{}}Obstructed\\ View\end{tabular}}} 
														& EKF\T 		& 66.27\%  								& 81.83\% 								& 78.30\% \\
		\multicolumn{1}{c|}{}    						& TD\T  		&  69.80\%  							& 86.01\% 								&  82.48\% \\
		\multicolumn{1}{c|}{}       					& RAPS\T 		& \textbf{82.09\%}  					& \textbf{91.63\%} 						& \textbf{97.52\%} \\ \hline
		\multicolumn{1}{c|}{\multirow{3}{*}{Open sky}}  & EKF\T 		& 96.03\%  								& \textbf{98.91}\% 						& \textbf{99.13}\% \\
		\multicolumn{1}{c|}{}     						& TD\T 			& \textbf{96.30}\%  					& \textbf{98.91}\%  					& 98.91\% \\
		\multicolumn{1}{c|}{}     						& RAPS\T 		&   92.61\%  							&  97.83\% 								& \textbf{99.13}\%  \\ \hline
	\end{tabular}
	\caption{Horizontal and vertical position error probabilities.}
	\label{fig:status_pr}
\end{table}

\section{Conclusion and Discussion}\label{sec:conclu}
This paper presents, implements, and demonstrates the diagonal specification \ac{RAPS} approach for outlier accommodation with a focus on \ac{GNSS} single frequency applications using  \ac{RTPPP} correction information. 
%Algorithms suitable for real-time computation are demonstrate.
\ac{EKF}, traditional \ac{TD}, and \ac{RAPS} approaches are evaluated and compared.
All estimators have the option to select from the same \ac{GNSS} pseudorange and Doppler measurements to use at each epoch. 
Each algorithm makes different choices resulting in different risk and performance. 
The experimental data set include  both open sky and obstructed view conditions.
The results demonstrate that all three approaches achieve performance that exceeds the SAE specification when averaged over the entire data set.
In challenging time intervals with obstructed views or heightened multipath effects, the \ac{RAPS} approach, using diagonal performance specification constraints, outperforms the rest. 
It shows improvements of 6-12\% over \ac{TD} and 10-19\% over \ac{EKF}.

Several avenues can be explored in future research. 
The extension to carrier-phase measurements would be valuable for enhanced precision. 
%Incorporating carrier-phase measurements aligns with the \ac{PPP} \ac{RTK} technique. 
However, leveraging carrier-phase measurements in an single frequency context is challenging to to inaccuracy in the ionospheric PPP products, whose accuracy may be enhanced with the launch of Stage 3 of \ac{IGS} schedule for the near future \cite{igs2020ssr}.
Multi-frequency scenarios are also of interest for addressing the ionospheric delay to make carrier phase feasible in life-critical applications.

\section{Acknowledgment}
The authors gratefully acknowledge the UCR KA Endowment and the US DOT CARNATIONS Center, each of which has provided partial funding for this research. 

The ideas reported herein, and any errors or omissions, are the responsibility of the authors and do not reflect the opinions of the sponsors.

\bibliographystyle{biblio/IEEEtran}
\bibliography{biblio/IEEEabrv,biblio/References.bib}

\end{document}